\documentclass[pra,showpacs,twocolumn,superscriptaddress]{revtex4}

\newcommand{\ket}[1]{|{#1}\rangle}
\newcommand{\bra}[1]{\langle{#1}|}

\newcommand{\Melement}[3]{\left\langle #1 \right| #3 \left| #2 \right\rangle } 

\usepackage{amsfonts}
\usepackage{amsmath}
\usepackage{graphicx}
\usepackage{amssymb}

\begin{document}
\title{Quantized vortices with density-dependent synthetic gauge fields}

\author{Salvatore Butera}
\author{Manuel Valiente}
\author{Patrik \"Ohberg}
\affiliation{SUPA, Institute of Photonics and Quantum Sciences, Heriot-Watt University, Edinburgh EH14 4AS, United Kingdom}
\begin{abstract}
We consider a two-dimensional weakly interacting ultracold Bose gas whose constituents are two-level atoms. We study the effects of a synthetic density-dependent gauge field that arises from laser-matter coupling in the adiabatic limit with a laser configuration such that the single-particle zeroth-order vector potential corresponds to a constant synthetic magnetic field.
We find a new exotic type of current non-linearity in the Gross-Pitaevskii equation which affects the dynamics of the order parameter of the condensate. We investigate the rotational properties of this system {in the Thomas-Fermi limit}, focusing in particular on the physical conditions that make the {existence} of a quantized vortex in the system energetically favourable with respect to the non-rotating solution.
We point out that two different physical interpretations can be given to this new non linearity: firstly it can be seen as a local modification of the mean field coupling constant, whose value depends on the angular momentum of the condensate. Secondly, it can be interpreted as a density modulated angular velocity given to the cloud. Looking at the problem from both of these viewpoints, we show that the effect of the new nonlinearity is to induce a rotation to the condensate, where the transition from non-rotating to rotating states depends on the density of the cloud.
\end{abstract}
\maketitle

\section{Introduction}
Since its theoretical prediction and first experimental realization in 1995 \cite{Anderson1995,Davis1995a}, Bose-Einstein condensation in atomic gases has drawn the attention of the scientific community. One of the most intriguing phenomena that have been studied since the realization of these macroscopic quantum fluids, and in which the signature of quantum physics is evident, is represented by the quantized circulation of the velocity field \cite{Matthews1999a,Madison2000,Abo-Shaeer2001,lin_2009b}. The feature that differentiates most a quantum fluid from a classical one, is in the way it responds to rotation. In contrast to a classical fluid, which rotates as a rigid body, a Bose-Einstein condensate rotates by forming quantized vortices.

The physics of rotating superfluids and their vortex structure is well studied \cite{Fetter2009}, ranging from early studies of superfluid Helium \cite{kapitza_1938,allen_1938} to vortices in strongly interacting quantum gases \cite{Zwierlein2003}. Moreover, strongly correlated phases in the limit of high vortex numbers, which can be regarded as bosonic quantum Hall states, can also appear in rotating condensates \cite{cooper_2008}. Rotating a condensate does however have its limitations. It is only equivalent to a uniform magnetic field, and it is not always an easy task to control and rotate a small number of particles which is needed if one wants to obtain a large flux per particle and study quantum Hall physics. For sufficiently rapid rotation, or strength of the synthetic magnetic field, one finds that the ground state of the Bose-Einstein condensate will correspond to a vortex state. 
The experimental progress, however, in realising rapidly rotating superfluids with ultracold atoms came by taking a different route. It was the emulation of orbital magnetism for neutral atoms taking advantage of laser-matter coupling that made it possible to study magnetic effects without physically rotating the gas \cite{dalibard_2011,goldman_2013}. This has opened up a plethora of studies, both in terms of experiments and theory, which cover phenomena such as quantised vortices \cite{lin_2009b}, spin-orbit coupling\cite{Lin2011,Wang2012,Cheuk2012,Zhang2012PRL} and various concepts from mathematical physics and high energy physics \cite{Bermudez:2010,Endres2012,Kapit2011,Bissbort2011,Banerjee2013,ray-2014}.
In the standard scenario with synthetic gauge potentials, the orbital magnetism stems from the interaction between light and matter. This means that the emerging gauge potentials are static, and are only given by the parameters of the incident lasers. In other words, there is no back-action between the gauge field and the matter field. A non-linear gauge potential \cite{edmonds_2013a} on the other hand, would change this fact, and result in some intriguing dynamics \cite{edmonds_2014}.

In this paper we will address one basic question: how are the ground state properties and vortex solutions affected by the presence of a nonlinear gauge potential which is proportional to the density of the condensate? We will consider an atomic Bose-Einstein condensate subject to a laser field configuration which gives rise to a synthetic vector potential, whose zeroth-order contribution {with respect to the ratio between the atom-atom and atom-field interaction}, emulates a constant magnetic field. However, in a mean field picture, detunings from the atomic resonance can be induced by collisions, which results in an unconventional current nonlinearity \cite{edmonds_2013a,greschner_2013,zheng_2014}, and corresponds to the semiclassical, or mean-field limit of an interacting gauge theory \cite{aglietti_1996}. 

In two dimensions, we show that the current non-linearity gives rise to a non-trivial non-linear term, proportional to the angular momentum of the condensate. It is different from the standard nonlinearity appearing in the Gross-Pitaevskii (GP) equation, which affects the rotational properties of the condensate in an indirect way by modifying the structure of a vortex state, in particular the size of the vortex core which is of the order of the healing length of the condensate, and the overall width of the cloud. We will show that the new nonlinearity mentioned above has direct effects on the vorticity of the system, because of its explicit dependence on the angular momentum.  We address in particular the problem of the on-set of vorticity in the condensate due to the current non-linearity. As in the standard case of a constant magnetic field,  the effect of the density dependent gauge potential is to induce a rotation to the condensate, with the main difference that the onset of vorticity now depends on the number of particles in the condensate: the higher the density is, the more likely it is to have vortices. 

After introducing the mechanism by which an artificial gauge field acting on the  condensate is generated, and deriving the most general expressions for the synthetic scalar and vector potentials, we introduce the conditions under which a constant magnetic field is generated. We derive the corresponding form of the generalised Gross-Pitaevskii equation, showing the appearance of a current nonlinearity. Finally we study {analytically} the conditions under which a vortex state appears as the ground state for the system {in the Thomas-Fermi limit}, comparing the result with the standard case of a rotating condensate.

\section{Atoms in artificial gauge fields}
We consider a Bose gas of two-level atoms and we model the short-range interactions by a zero-range pseudo-potential. In addition, we include a coupling between the two levels, mediated by an external laser field. In the rotating wave approximation (RWA), {which is valid when the frequency of the applied electric field is close to the atomic resonance, so that the terms in the Hamiltonian describing the atom-light interaction which are rapidly oscillating can be averaged to zero, in contrast to the slowly oscillating ones \cite{loudon_2000}}, the microscopic $N$-body Hamiltonian of the system is given by
\begin{eqnarray}
	H&=&\sum_{n=1}^{N}{\left(\frac{\mathbf{p}_n^2}{2m}+U_{af}(\mathbf{r}_n)+V(\mathbf{r}_n)\right)\otimes\mathbb{I}_{\mathcal{H}\backslash n}}\nonumber \\&& +\sum_{n<\ell}^N{\nu_{n,\ell}\otimes\mathbb{I}_{\mathcal{H}\backslash\{n,\ell\}}},
\label{Micr_Hamil}
\end{eqnarray}
where the first term is the non-interacting Hamiltonian, with $V(\mathbf{r})$ the external trapping potential, $\mathbf{p}_n$ the quantum momentum operator of the n-th particle, and $m$ its mass. The identity matrices $\mathbb{I}_{\mathcal{H}\backslash\{n,\ell,...\}}$ act on the subspace excluding the particles $n,\ell,...$. Finally, the coupling $U_{af}(\mathbf{r})$ between the two atomic levels $\ket{1}$ and $\ket{2}$ is given by
\begin{equation}
	U_{af}(\mathbf{r})=\frac{\hbar\Gamma}{2} \begin{pmatrix}
		\cos\theta(\mathbf{r}) & e^{-i\phi(\mathbf{r})}\sin\theta(\mathbf{r})\\
	 e^{i\phi(\mathbf{r})}\sin\theta(\mathbf{r}) & -\cos\theta(\mathbf{r})
	\end{pmatrix}.
\label{Uaf^n}
\end{equation}
The light-matter interaction, eq. (\ref{Uaf^n}), is fully characterized by three parameters, namely the generalized Rabi frequency $\Gamma$, the mixing angle $\theta(\mathbf{r})$ and the laser's phase $\phi(\mathbf{r})$ at the atom's position $\mathbf{r}$. {These parameters, useful in order to make evident the unitary property of the operator, are related to the laser detuning $\Delta$ and the Rabi frequency $\kappa$ as $\Delta=\Gamma\cos\theta$ and $\kappa=\Gamma\sin\theta$. The detuning $\Delta$ is defined as the difference between the incident laser frequency and the frequency of the transition between the two atomic levels. The Rabi frequency $\kappa$, characterizing the strength of the atom-laser coupling, is instead defined in terms of the matrix element $\mathbf{d}_{12}=\Melement{1}{2}{\mathbf{d}}$ of the atomic electric dipole operator between the two internal states, and the amplitude  $\mathbf{E}$ of the of the electric field, as $\kappa=\mathbf{d}_{12}\cdot \mathbf{E}/\hbar$ \cite{loudon_2000}}.\\
The second term in the Hamiltonian \eqref{Micr_Hamil} is the pairwise interaction between the particles, and has the diagonal form $\nu_{n,\ell}=\text{diag}\left[g_{11},g_{12},g_{12},g_{22}\right]\,\delta\left(\mathbf{r}_n-\mathbf{r}_\ell\right)$. The coupling constants are given by $g_{ij}={4\pi\hbar^2 a_{ij}}/{m}$, where $a_{ij}$ are the scattering lengths for the three different collision channels.

In what follows we shall work in the weakly interacting limit, {$\rho_i a_{ij}^3\ll 1$ (with $i,j=1,2$),} and we consider the meanfield variational ansatz for the many-body wave function of the system: $\boldsymbol{\Psi}\left(\mathbf{r}_1,\mathbf{r}_2,...\mathbf{r}_N\right)=\prod_{i=1}^N\phi\left(\mathbf{r}_i\right)$, that is the symmetrized product of the single particle spinor wave function $\boldsymbol{\phi}(\mathbf{r})$, satisfying the normalization condition $\int{d^3 \mathbf{r}\,\boldsymbol{\phi}^\dag \boldsymbol{\phi}}=1$. We introduce the Lagrangian of the system, which reads as
\begin{equation}
	L=\int{\prod_{i=1}^N d^3 \mathbf{r}_i}\left[\boldsymbol{\Psi}^\dag\left(i\hbar\partial_t-H\right)\boldsymbol{\Psi}\right].
\label{Lagrangian_many}
\end{equation}
Substituting in eq.\eqref{Lagrangian_many} the expression given above for the many-body wave function, we obtain the corresponding Lagrangian in terms of the single particle wave function
\begin{equation}
	L_{MF}=\int{d^3 \mathbf{r}}\left[\boldsymbol{\psi}^\dag\left(i\hbar\partial_t-H_{MF}\right)\boldsymbol{\psi}\right]
\label{Lagrangian_single}
\end{equation}
where $\boldsymbol{\psi}(\mathbf{r})=\sqrt{N}\boldsymbol{\phi}(\mathbf{r})$ is the order parameter of the system (or wave function of the condensate), and we defined the single particle mean field Hamiltonian $H_{MF}$ as
\begin{equation}
	H_{MF}=\frac{\mathbf{p}^2}{2m}\otimes\mathbb{I}+U_{af}(\mathbf{r})+V_{aa}+V\left(\mathbf{r}\right)
\label{MF_Hamil}
\end{equation}
with $\mathbb{I}$ the $2\times 2$ identity operator in the space of the atomic internal degrees of freedom. In eq.\eqref{MF_Hamil} $V_{aa}$ describes the mean field collisional effects, and is given by
\begin{equation}
	V_{aa}=\frac{1}{2} \begin{pmatrix}
		\nu_1 & 0\\
	 0 & \nu_2
	\end{pmatrix}
\label{Vaa}
\end{equation}
where
\begin{align}
	\nu_1&=g_{11}\rho_1+g_{12}\rho_2\\
	\nu_2&=g_{12}\rho_1+g_{22}\rho_2
\label{nu1_nu2}
\end{align}
and  $\rho_{i}=\left|\psi_i\right|^2$ is the density of atoms in level $\ket{i}$, $i=1,2$ {(with $\psi_i$ the corresponding component of the order parameter).}

In the weakly interacting limit, the coupling strength $\hbar\Gamma$ between the ground and the excited states is typically much larger than the mean field energy shifts. This allows us to diagonalize the Hamiltonian in eq. \eqref{MF_Hamil} by treating the mean field interaction term as a small perturbation to the atom-field interaction. To order $\mathcal{O}(\nu_{d}/\hbar\Gamma)$, the eigenstates of eq. \eqref{MF_Hamil} take the form
\begin{equation}
	\ket{\chi_\pm}=\ket{\chi_\pm^{(0)}}\pm\frac{\nu_d}{\hbar\Gamma}\ket{\chi_\mp^{(0)}}.\label{IntDress}
\end{equation}
{where $\nu_d$ is the pertubative parameter, equal to
\begin{equation}
	\nu_d=\frac{1}{2}\sin\frac{\theta}{2}\cos\frac{\theta}{2}\left(\nu_1 -\nu_2 \right)
\label{nu_pm}
\end{equation}
and $\ket{\chi_\pm^{(0)}}$ are the unperturbed dressed states, having the form
\begin{align}
	\ket{\chi_{+}^{(0)}}&=\cos\frac{\theta}{2}\ket{1}+e^{i\phi}\sin\frac{\theta}{2}\ket{2}\nonumber\\
	\ket{\chi_{-}^{(0)}}&=\sin\frac{\theta}{2}\ket{1}-e^{i\phi}\cos\frac{\theta}{2}\ket{2}
\label{chi_pm}
\end{align}
}
The condensate wave function $\ket{\psi(\mathbf{r},t)}$ can then be expanded in the interacting dressed state basis, eq. (\ref{IntDress}), as $\ket{\psi(\mathbf{r},t)}=\sum_{i=\{+,-\}}{}\psi_i(\mathbf{r},t)\ket{\chi_i}$.
{In order to capture the dynamics of the $\pm$ component of the condensate we use the adiabatic assumption, according to which $\psi_\mp(\mathbf{r},t)\equiv 0$, and we consider the projection of the mean field Lagrangian in eq.\eqref{Lagrangian_single} onto the subspace spanned by the corresponding $\left(\ket{\chi_\pm}\right)$ dressed state. We obtain then the mean field Lagrangian for the relevant condensate component that, to order $\mathcal{O}(\nu_{d}/\hbar\Gamma)$ reads as:
\begin{equation}
	L_\pm=\int{d^3 \mathbf{r}}\left[\psi_\pm^\dag\left(i\hbar\partial_t-H_\pm\right)\psi_\pm\right]
\label{Lagrangian_pm}
\end{equation}
where
\begin{equation}
	H_\pm=\frac{\left(\mathbf{p}-\mathbf{A}_\pm\right)^2}{2m}+W\pm\frac{\hbar\Gamma}{2}+\nu_\pm + V\left(\mathbf{r}\right).
\label{Hamil_pm}
\end{equation}
is the Hamiltonian describing the dynamics of the $\pm$ component of the condensate.
}
Above {we defined
\begin{align}
	\nu_+&=\frac{1}{2}\left(\nu_1 \cos^2\frac{\theta}{2}+\nu_2 \sin^2\frac{\theta}{2}\right)\nonumber\\
	\nu_-&=\frac{1}{2}\left(\nu_1 \sin^2\frac{\theta}{2}+\nu_2 \cos^2\frac{\theta}{2}\right)\nonumber\\
\label{nu_p_m}
\end{align}
} while $\mathbf{A}_\pm$ and $W$ are effective vector and scalar potentials, respectively. Exploiting the identity relation $\mathbb{I}=\ket{\chi_+}\bra{\chi_+}+\ket{\chi_-}\bra{\chi_-}$, it is easy to show that they get the form (see \cite{dalibard_2011,goldman_2013}):
\begin{equation}
\begin{split}
	\mathbf{A}_\pm&=-\Melement{\chi_\pm}{\chi_\pm}{\mathbf{p}}\\
	W&=\frac{\left|\Melement{\chi_+}{\chi_-}{\mathbf{p}}\right|^2}{2m}
\end{split}
\label{potentials_def}
\end{equation}
{The adiabatic approximation, used in order to derive the Hamiltonian in eq.\eqref{Hamil_pm}, is satisfied when the speeds of the particles are low enough that the time scale of their motion is much longer than the time scale given by the Rabi frequency. Since the energy splitting between the interacting dressed state is proportional to the Rabi frequency in fact (and equal to $\hbar\Gamma+(\nu_+-\nu_-)$), this condition prevents (or at least attenuates) the transition of the atoms from one state to the other.} Upon substitution of eqs. \eqref{chi_pm}, \eqref{nu_pm} and \eqref{IntDress} into eq.\eqref{potentials_def}, we obtain the following explicit forms for the vector and scalar potentials
\begin{eqnarray}
     \mathbf{A}_\pm&=&-\frac{\hbar}{2}(1\mp\cos\theta)\boldsymbol{\nabla}\phi+\frac{f_\pm}{8\Gamma}\rho_\pm\boldsymbol{\nabla}\phi \label{a1}\\
      2mW&=&\frac{\hbar^2}{4}(\boldsymbol{\nabla}\theta)^2+\hbar^2\frac{\sin^2\theta}{4}(\boldsymbol{\nabla}\phi)^2-\nonumber\\&&\frac{\hbar f}{8\Gamma}\rho_+\cos\theta(\boldsymbol{\nabla}\phi)^2-\nonumber\\
	&&\hbar\boldsymbol{\nabla}\theta\cdot\boldsymbol{\nabla}\left(\frac{\nu_d}{\Gamma}\right)
 \label{potentials_extended}
\end{eqnarray}
where we defined the coefficients $f_\pm$ as
{
\begin{align}
	f_+&=2\sin^2\theta\left[\cos^2\frac{\theta}{2}g_{11}+g_{12}\left(\sin^2\frac{\theta}{2}-\cos^2\frac{\theta}{2}\right)\right.\nonumber\\
	&\left.-g_{22}\sin^2\frac{\theta}{2}\right]\nonumber\\
	f_-&=-f_+(1\leftrightarrow 2)
\label{f_pm} 
\end{align}
}
and with $\rho_\pm\equiv |\psi_\pm|^2$ the density of atoms in the $\ket{\chi_\pm}$ state.
As we observe in eqs. \eqref{a1} and \eqref{potentials_extended}, the vector and the scalar potentials also depend on the particle density. In the following we analyse the important case in which, by properly engineered laser fields, the light-atom interaction induces an effective constant magnetic field, together with an extra nonlinear term that depends on the angular momentum of the condensate itself. 

\section{Physical assumptions and the equation of motion}

We choose a particular laser configuration, such that the density dependent terms in the gauge potentials lead to an exotic nonlinearity in the Gross-Pitaevskii equation which is proportional to the angular momentum of the condensate and we will investigate the rotational properties of such a system. With such setup, the lowest-order term in the vector potential corresponds to a constant magnetic field in the $z$-direction, whose effect on the physical properties of a condensate  have been widely studied in the literature and are nowadays well understood \cite{Fetter2009}.  It leads for instance to the nucleation of quantized vortices for sufficiently strong magnetic fields \cite{lin_2009b}, a situation that is  equivalent to considering a rotating condensate in the rotating frame when the magnetic field $q{\bf B}=2m\Omega {\bf \hat z}$, with $\Omega$ the rotation frequency {(and $q=1$  in our case)}. It is known in fact that once the angular velocity of the condensate (or the applied magnetic field) reaches a threshold value, a state with a vortex becomes energetically favourable with respect to a non-rotating one and, for strong magnetic fields, the Abrikosov triangular lattice of vortices is obtained. This type of ground state has been observed \cite{Abo-Shaeer2001} also in harmonically trapped clouds in the so called fast rotating regime, in which the centrifugal force nearly cancels the trapping potential. Based on these considerations, we will in the following address the question of how the density dependent gauge potential which, as said above, turns out to be proportional to the angular momentum of the condensate in our particular laser configuration, will affect the rotational properties of a condensate.

There are several ways of creating such non-trivial effective magnetic fields, and they typically depend on the details of the laser parameters. Here we choose to work in the symmetric gauge, where we obtain a constant magnetic field by making the zeroth-order term of the vector potential proportional to the radial distance from the origin of the reference frame, which we assume to coincide with the minimum of any potential trapping the cloud. This can be achieved by properly designing the spatial dependence of the three parameters that are involved, namely $\Delta$, $\kappa$ and the laser phase $\phi$, {in order to have $-\hbar/2\left(1\pm\cos\theta\right)\nabla\phi\propto r$. To this aim we choose to work with the $(+)$-component of the condensate, and} we choose a laser beam with a phase proportional to the polar angle $\varphi$ as $\phi=\ell\varphi$, where $\ell$ is the quantum number identifying the orbital angular momentum $\hbar \ell$ carried by the electromagnetic field. It can be shown that, by making the assumption $\kappa\ll\Delta$, the sought spatial dependence for the zeroth-order term is obtained, once we tune $\kappa$, $\Delta$ in such a way that $\kappa / \Delta\sim r$. It is worth noting that if we had chosen to work with the $(-)$-component of the condensate, the dynamics of the system would be the same provided we change the sign of the detuning, and we exchange $g_{11}\leftrightarrow g_{22}$.

The condition described above can be realised by designing the laser beam in such a way that the Rabi frequency is linear in $r$ in the plane in which we confine the cloud, and equal to $\kappa(r)=\kappa_0 r$ where $\kappa_0$ is a constant and also assuming that the detuning is constant in the same plane. These considerations are clear once we substitute the following relations in eq.\eqref{potentials_extended}:
\begin{align}
	\sin\theta&\simeq \kappa/\Delta \nonumber\\
	\cos\theta&\simeq1-\kappa^2/2\Delta^2 \nonumber\\
	\theta&\simeq\kappa/\Delta
\label{sincos}
\end{align}
that hold true in the limit $\kappa\ll\Delta$. In this limit, the perturbative condition $\nu_d\ll\hbar\Gamma$ reduces to $\nu_d\ll\hbar\Delta$, and the expressions for the potentials become
\begin{eqnarray}
	\mathbf{A}&\simeq&-\frac{\hbar}{4}\frac{\kappa^2}{\Delta^2}\boldsymbol{\nabla}\phi+\frac{\kappa^2}{\Delta^2}\frac{\gamma_{12}}{\Delta}\rho\boldsymbol{\nabla}\phi
	\\
	2mW &\simeq& \frac{\hbar^2}{4}\left(\boldsymbol{\nabla}\frac{\kappa}{\Delta}\right)^2+\frac{\hbar^2}{4}\frac{\kappa^2}{\Delta^2}(\boldsymbol{\nabla}\phi)^2-\hbar\frac{\kappa^2}{\Delta^2}\frac{\gamma_{12}}{\Delta}\rho\left(\boldsymbol{\nabla}\phi\right)^2\nonumber\\
	&&-\hbar\frac{\gamma_{12}}{\Delta}\left(\boldsymbol{\nabla}\frac{\kappa}{\Delta}\right)\cdot\left[\left(\boldsymbol{\nabla}\frac{\kappa}{\Delta}\right)\rho +\frac{\kappa}{\Delta}\boldsymbol{\nabla}\rho\right]
\label{potentials_approx}
\end{eqnarray}
together with 
\begin{eqnarray}
		\hbar\frac{\Gamma}{2}&\simeq&\frac{\hbar}{2}\Delta\left(1+\frac{\kappa^2}{2\Delta^2}\right)\\
		\nu_+&\simeq &\frac{g_{11}}{2}\rho - \gamma_{12}\frac{\kappa^2}{\Delta^2}\rho	
\label{other_terms_approx}
\end{eqnarray}
in which we defined $\gamma_{12}=\left(g_{11}-g_{12}\right)/4$, and we dropped for brevity the (+) subscript, since from now on we are dealing just with the atoms in their $\ket{\chi_+}$ internal eigenstate.


To justify the assumption we made for the ratio $\kappa/\Delta\propto r$, and in particular about how it is restricted by the condition $\kappa\ll\Delta$,
we define $\alpha\gg 1$ such that ${\kappa}/{\Delta}={\kappa_0\,r}/{\Delta}\leq {1}/{\alpha}$. This condition is fulfilled in the region $r\leq {\Delta}/{\kappa_0 \alpha}=R$ that also defines the region of validity of our model, and can be modified by properly tuning the constants $\Delta$ and $\kappa_0$.

In order to arrive at a Gross-Pitaevskii equation for the order parameter of the condensate, we minimize the action 

{
\begin{equation}
	S=\int{dt\,L_+}.
\label{action}
\end{equation}
}
{with respect to $\psi^*$.} This leads to the equation of motion
\begin{eqnarray}
i\hbar\frac{\partial\psi}{\partial t}&=&\bigg[\frac{\left(\mathbf{p}-\mathbf{A}\right)^2}{2m}+V\left(\mathbf{r}\right)+W+\frac{\hbar\kappa}{2}+\nu_+
\nonumber\\&&
+\left(\mathbf{a}_1\cdot\mathbf{j}_1+\mathbf{a}_2\cdot\mathbf{j}_2\right)
+\frac{\hbar}{2m}\frac{\gamma_{12}}{\Delta}\rho\frac{\kappa}{\Delta}\left(\boldsymbol{\nabla}^2\frac{\kappa}{\Delta}\right)\nonumber\\&&+\frac{g_{11}}{2}\rho+\gamma_{12}\rho\left(\frac{\kappa}{\Delta}\right)^2\bigg]\psi
\label{GP_full}
\end{eqnarray}
in which the nonlinear current $\mathbf{j}_1$ and the $\mathbf{j}_2$ term defined below appear, along with the coupling vector fields $\mathbf{a}_1$ and $\mathbf{a}_2$, with the form
\begin{eqnarray}
\mathbf{a}_1&=&\frac{\kappa^2}{\Delta^2}\frac{\gamma_{12}}{\Delta}\boldsymbol{\nabla}\phi\label{currents1}\\
\mathbf{j}_1&=&\frac{\hbar}{2im}\left[\psi\left(\boldsymbol{\nabla}-i\frac{\boldsymbol{\nabla}\phi}{2}\right)\psi^*-\psi^*\left(\boldsymbol{\nabla}+i\frac{\boldsymbol{\nabla}\phi}{2}\right)\psi\right]\label{currents2}\\
\mathbf{a}_2&=&\frac{\kappa}{\Delta}\frac{\gamma_{12}}{\Delta}\left(\boldsymbol{\nabla}\frac{\kappa}{\Delta}\right)\label{currents3}\\
\mathbf{j}_2&=&\frac{\hbar}{2im}\left[\psi^*\boldsymbol{\nabla}\psi+\psi\boldsymbol{\nabla}\psi^*\right]. \label{currents4}
\end{eqnarray}
Using the explicit expressions, $\frac{\kappa}{\Delta}=\frac{\kappa_0}{\Delta}r$ and $\phi=\ell\theta$, the Gross-Pitaevskii equation reads as
\begin{eqnarray}
i\hbar\frac{\partial\psi}{\partial t}&=&\left[-\frac{\hbar^2}{2m}\boldsymbol{\nabla}^2+\frac{\hbar\ell}{4m}\left(\frac{\kappa_0}{\Delta}\right)^2 L_z +V\left(\mathbf{r}\right)\right]\psi\nonumber\\
&&-\rho\bigg[\left(\frac{\kappa_0}{\Delta}\right)^2\frac{\gamma_{12}}{\Delta m}\left(2\ell L_z+\frac{\hbar}{2}(2\ell^2+1)\right)\nonumber\\&&-\left(g_{11}+2\gamma_{12}\frac{\kappa^2}{\Delta^2}\right)\bigg]\psi
\label{GP_extended}
\end{eqnarray}
where $L_z=i\hbar\left(y\frac{\partial}{\partial x}-x\frac{\partial}{\partial y}\right)$ is the orbital angular momentum operator.
As expected, the linear term in eq.\eqref{GP_extended} is identical to the Hamiltonian describing the motion of a charged particle in a constant magnetic field (or of a rotating condensate). The relation between the effective magnetic field $B$, the angular velocity $\Omega$ and the parameter of our system, is now given by
\begin{equation}
	B=2m\Omega=\frac{\hbar\ell}{2}\left(\frac{\kappa_0}{\Delta}\right)^2. 
\label{correspondences}
\end{equation}
The first bracket in eq. \eqref{GP_extended} describes the dynamics of a particle, subjected to the effective magnetic field \eqref{correspondences} whose intensity is proportional to the orbital angular momentum $\hbar\ell$ carried by the laser beam, and confined by the potential $V\left(\mathbf{r}\right)$, while the second bracket is the resulting nonlinear term of first order in the atom-atom interaction. In the latter, the second term contains the standard mean field term proportional to the density $g_{11}\rho$. We will drop the last term in what follows, since it is of order $\mathcal{O}(\kappa/\Delta)^2$, and corresponds to a scalar interaction much smaller than $g$ in the centre of the trap. The goal is to study the role of the nonlinear term proportional to the orbital angular momentum $L_z$, since the remaining terms only modify the effective mean field coupling constant. In order to isolate the effect of the density dependent gauge potential we envisage a situation where we counteract the linear zeroth order gauge potential by rotating the system at the frequency $\Omega=-\frac{\hbar\ell}{4m}\left(\frac{\kappa_0}{\Delta}\right)^2$, along with the laser apparatus. 
This might be  experimentally challenging \cite{spielman_2009}, but it allows us to focus on  the effect of the density dependent gauge potential, from which novel phenomena \cite{edmonds_2014} is expected to arise. The resulting Gross-Pitaevskii equation, whose study  will be the object of the next section, takes finally the form
\begin{equation}
i\hbar\frac{\partial\psi}{\partial t}=\left[-\frac{\hbar^2}{2m}\boldsymbol{\nabla}^2-C\rho L_z+V\left(\mathbf{r}\right)+g\rho\right] \psi
\label{GP_final}
\end{equation}
where we have defined $C=\frac{2\ell}{m}\left(\frac{\kappa_0}{\Delta}\right)^2\frac{\gamma_{12}}{\Delta}$.

\section{The vortex state}
In the previous section we derived the Gross-Pitaevskii equation for a condensate with an angular momentum non-linearity. We will next focus on the physical conditions for the nucleation of vortices in the condensate. In order to extract any new phenomena deriving from the nonlinearity, we will consider the case of a state containing a single vortex state carrying one quantum of circulation. 

An interesting point that is worth emphasizing is that there are two different ways in which we can look at the nonlinear angular momentum term. On one hand, it can be seen as a modification of the mean field coupling constant, which depends on the local angular momentum of the condensate. This results in a breaking of the rotational  invariance of the system, enhancing or weakening the effective interaction felt by the atoms, depending on the direction of rotation of the condensate. On the other hand, the nonlinearity can also be understood as a density modulated angular velocity of the cloud. In order to see to what extent the rotational properties of the condensate are modified, we start the discussion considering this latter interpretation, looking for the critical value of the $C$ parameter in eq. \eqref{GP_final}, that makes a vortex  state of the condensate energetically favourable with respect to its non-rotating solution. {To this aim we consider a BEC in a highly anisotropic trap, such that its dynamics in the transversal direction is frozen and the system is quasi-two-dimensional, and confined in its plane by an isotropic harmonic potential of frequency $\omega_0$. We assume the condensate uniform in the $z$ direction, with thickness $Z$. The original order parameter can then be rescaled as $\psi(\mathbf{r})/\sqrt{Z}$, where $\psi(\mathbf{r})$ is now two-dimensional, and normalized in such a way that $\int{d^2 r |\psi|^2}=N$. All the previous expressions for the Hamiltonian $H_\pm$ and the potentials $\mathbf{A}_\pm$, $W$ hold true, with the meanfield coupling constants and the $C$ parameter rescaled as $g\rightarrow g/Z$ and $C\rightarrow C/Z$.} We consider the condensate composed by a number of atoms sufficiently large so that its physical properties are well described working in the Thomas-Fermi limit \cite{pethick-book}.

In this regime, {the width $R$ of the condensate is much larger than the healing length $\xi$ so that, because of the large $r$ behaviour of the azimuthal velocity field,} the extra energy associated with the presence of a vortex (compared to the energy of the same non-rotating system) can be estimated {by considering just the corresponding kinetic term}  \cite{pethick-book} giving $E_V=\pi \rho(0) \frac{\hbar^2}{m} \log\left(0.888 R/\xi\right)$, where {$\rho(0)$ is the density of the non-rotating condensate at the center of the trap}. In the standard situation in which only the linear rotational term $-\Omega L_z \psi$ is in the Gross-Pitaevskii equation, the vortex state becomes energetically favourable when the energy $E_V'=E_V-\Omega \langle L_z \rangle$ of the excitation in the reference frame co-rotating with the condensate is less than the energy corresponding to the non rotating cloud, with $\langle L_z\rangle$ the value of the angular momentum in the single vortex state. This condition leads to the critical angular velocity
\begin{equation}
	\Omega_{cr}=\frac{E_V}{\langle L_z\rangle}=\frac{2\hbar}{m R^2}\log\left(0.888 R/\xi\right)
\label{Critical_linear}
\end{equation}
where $\langle L_z\rangle=\int{d^2 r\psi^*L_z \psi}=\frac{1}{2}\rho(0)\pi\hbar R^2$, having used the expression $\rho(r)=\rho(0)\left(1-r^2/R^2\right)$ for the Thomas-Fermi density profile.  Motivated by this, we note that, with the nonlinear term $-C\rho(\mathbf{r}) L_z \psi$ in the Gross-Pitaevskii equation, we can estimate the critical value $C_{cr}$ in the same fashion, obtaining:
\begin{equation}
	C_{cr}=\frac{3}{2}\frac{\Omega_{\text{cr}}}{\rho(0)}
\label{Critical_nonlinear}
\end{equation}
having used  $\langle\rho L_z\rangle=\int{d^2 r\rho(\mathbf{r})\psi^*L_z \psi}=\frac{\pi}{3}\hbar\rho^2(0) R^2$. Apart from a constant factor then, the critical value of $C$ is given by $C_{cr}\sim\frac{\Omega_{cr}}{\rho(0)}$. 
The same result can be obtained by looking at the new nonlinearity as a modification of the mean field coupling constant. This effect is clear once we substitute the ansatz $\psi(\mathbf{r})=f(r) e^{i\varphi}$ for the order parameter into eq.\eqref{GP_final}, with the characteristics of a vortex state with angular momentum $\hbar$ {per particle}. In this way, we get the GP equation for the modulus $f(r)$ of the wave function, with the mean field coupling constant changing as $g\to g'=g-\hbar C$. The effective coupling constant then is stronger or weaker depending on the combinations of the scattering length differences and the orbital angular momentum of the incident laser beam, which can contribute both with a positive or negative sign.  

We then see that the nucleation of a vortex changes abruptly the properties of the condensate since it changes the mean field interaction strength, which also affects physical quantities such as the healing length and therefore also the shape and size of the vortex core. 
Moreover, the modification of such interaction strength shifts the energy of the system at hand, for which the energy of the non-rotating state can be taken as a reference value, and in respect to which all the excited states of the condensate are defined.

In the case $g'<g$, the abrupt drop of this background energy could make the vortex solution energetically favourable compared to the non rotating one. This happens when this energy  shift is larger than the energy $E_V$ of the vortex,  see Fig.1(a).

\begin{figure}[]
\centering
\includegraphics[width=1\linewidth]{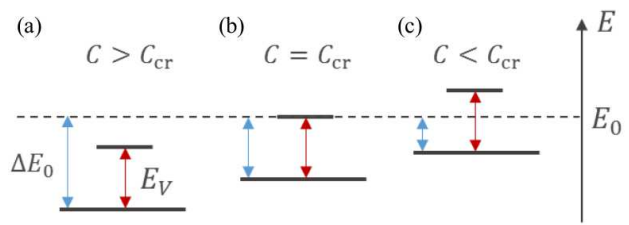}
\caption{Illustration of the conditions for the onset of vorticity in the condensate, when the energy of the vortex state is lowered by the effect of the laser field incident on the condensate itself. The solid long line represents the reference non rotating state energy and the solid short one the energy of the vortex state. (a) When the redefinition of the meanfield coupling constant $g'\rightarrow g-\hbar C$, due to the presence of the vortex in the superfluid, induces a shift $\Delta E_0$ of the non-rotating state energy $E_0$, which is larger than the energy $E_V$ of the vortex itself, then the rotating state becomes energetically favourable compared to the non rotating one. (b) The critical condition happens when $E_V$ equates the energy shift of the non rotating state. In the case (c), the energy of the vortex state is lowered, but not enough to induce its nucleation.}
\label{fig:spiral}
\end{figure}

In the TF limit, the total number of particles is related to the chemical potential by the equation
\begin{equation}
	N=\frac{\pi}{U_0}\frac{\mu^2}{m\omega_0^2}
\label{NumPart_ChemPot}
\end{equation}
having indicated with $U_0$ the generic value of the mean field coupling constant. Solving eq. \eqref{NumPart_ChemPot} for $\mu$ we get
\begin{equation}
	\mu=\left(\frac{m \omega_0^2 U_0}{\pi}\right)^{1/2}N^{1/2}
\label{ChemPot_NumPart}
\end{equation}
and since by definition $\mu=\frac{\partial E}{\partial N}$ we get, by using  eq.\eqref{ChemPot_NumPart}
\begin{equation}
	E_0(N)=\int_0^N{dN' \mu(N')}=\frac{2}{3}\mu N.
\label{En_N}
\end{equation}
The energy difference $\Delta E_0$ between the two non-rotating states with values $U'_0=U_0-\Delta U_0$ and $U_0$ for the mean field coupling constant, is equal to:
\begin{equation}
	\Delta E_0=E_0(U_0)-E_0(U_0-\Delta U_0)=\frac{\partial E_0\left(U_0\right)}{\partial U_0}\Delta U_0
\label{EnDiff_1}
\end{equation}
for $\Delta U_0\ll U_0$. Differentiating eq.\eqref{ChemPot_NumPart} with respect to $U_0$, we have $\partial\mu/\partial U_0=1/2\, \mu/U_0$ so that:
\begin{equation}
	\Delta E_0=\frac{1}{3}\mu N\frac{\Delta U_0}{U_0}.
\label{EnDiff_2}
\end{equation}
This energy shift has to be compared with the energy of the vortex state $E_v\left(U_0-\Delta U_0\right)=E_v\left(U_0\right)-\left[\partial E_v\left(U_0\right)/\partial U_0\right],\Delta U_0$. Since $\rho(0)=\mu/U_0$, we get $\partial\rho(0)/\partial U_0=-1/2\,\mu/U_0^2$, with $\left[\partial E_v(U_0)/\partial U_0\right] \,\Delta U_0=-1/2\,(\mu/U_0)\,(\Delta U_0/U_0)$ being a higher order correction to the vortex energy $E_v\left(U_0\right)$, and can be neglected. Equating $E_v\left(g\right)$ with the value $\Delta E_0$ calculated for $\Delta U_0=\hbar C$, that in ordinary cold atoms experiments fulfills the condition $\hbar C \ll g$, the critical value for the $C$ parameter takes the form
\begin{equation}
	C_{\text{cr}}=3\frac{\Omega_{\text{cr}}}{\rho(0)}
\label{Critical_nonlinear_2}
\end{equation}
that is the same result as in eq. \eqref{Critical_nonlinear}, up to a multiplicative factor of 2. This discrepancy can be ascribed to the fact that the result in eq.\eqref{Critical_nonlinear} represents an estimation for $C_{\text{cr}}$, since it is obtained by making an equivalence between the system at hand and a rotating condensate, that actually are two different physical situations.

This result tells us that the effect of the synthetic gauge potential is in fact to induce a rotation in the condensate, self-boosted by the total number of atoms composing the condensate itself. The higher this number is, the easier it is to induce the nucleation of vortices and so to put the cloud into rotation. Remembering the definition of the parameter $C$ in terms of the physical quantities characterizing the system, we obtain the critical value
\begin{equation}
	\rho_{\text{cr}}(0)=\frac{3\hbar\Delta}{\ell\gamma_{12}R^2}\left(\frac{\Delta}{\kappa_0}\right)^2 \log\left(0.888 R/\xi\right)
\label{ell_crit}
\end{equation}
of the particles number for which a vortex state is favoured. From the expression in eq. (\ref{ell_crit}) we see that there is significant flexibility in reaching the required critical combination of parameter values, due to the the fact that the orbital angular momentum of the light beam $\ell$, the scattering length difference $\gamma_{12}$, and also to some extent the density of the cloud, can all be relatively easily changed in an experiment. For example, we can choose the scattering length difference $a_{11}-a_{12}$ to be $100\,\text{nm}$ by using Feshbach resonances if necessary, a Rabi frequency of the order of $10^4\,\text{Hz}$ at the border of the cloud, and a value for the homogeneous laser detuning from the atomic resonance such that ${\kappa_0 R}/{\Delta}=0.1$, satisfying the assumption $\kappa\ll\Delta$ throughout the condensate (with $\alpha=10$ at the border of the cloud). A metastable Helium gas could be particularly useful in our case because of the small mass of the particles ($\sim 10^{-27}\text{kg}$). Considering an angular momentum carried by the laser beam characterized by the quantum number $\ell\simeq 100$ \cite{oam-vienna}, we obtain a critical value for the particle density at the centre of the trap $\rho_{cr}(0)$ of the order of $10^{14}$ atoms/$\text{cm}^3$, for which the vortex state is energetically favourable compared to the non rotating one. This density can be reached in current experiments  for which the condition of weakly interacting atoms $\rho a^3\ll 1$ is still fulfilled.

The result in eq.\eqref{ell_crit}, obtained for the case of a vortex state carrying a single quantum of circulation, can be easily generalised to the case of a vortex with higher angular momentum. Similar results are also expected for states with more than one vortex.

\section{Conclusions}

In this paper we have studied the {energetic properties of a singly quantized vortex state in} a Bose-Einstein condensate which is subject to a density dependent gauge potential. Nontrivial nonlinear dynamics emerges through the collisionally induced detuning for the light-matter coupling, which manifests itself as a nonlinearity where the density times the angular momentum of the condensate appears. We have chosen to study one particular setup using an incident laser beam with an orbital angular momentum, which couples the two internal levels of the atoms. This is only one possibility; there are many other setups which can be used which do not explicitly rely on a symmetric gauge \cite{dalibard_2011,goldman_2013} but will provide a uniform magnetic field for the zeroth order term. Nevertheless, the main conclusion still remains, the onset of a rotating state via the effective magnetic field, will depend on the density of the condensate in combination with other parameters, in stark contrast to the standard situation of vortex creation. Apart from the fundamental interest in rotating superfluids, the results presented here also shine light on the unprecedented flexibility in creating novel types of nonlinear dynamics for the Bose-Einstein condensate. The current non-linearity used in this work opens up many new questions regarding the superfluid dynamics. For instance, drag forces and the vortex dynamics including vortex interactions may well be dramatically altered due to the current non-linearity.

\section{Acknowledgements}

We acknowledge fruitful discussions with Luis Santos. S.B acknowledges support from the EPSRC CM-DTC Grant No. EP/G03673X/1, M.V and P.\"O acknowledge support from EPSRC EP/J001392/1.

\bibliography{vortex_bib}

\end{document}